\documentclass[showpacs,preprintnumbers,amsmath,amssymb,aps]{revtex4}

\begin{document}

\title{Anisotropic induced gravity and inflationary universe}
\author{
W. F. Kao\thanks{%
gore@mail.nctu.edu.tw} \\
Institute of Physics, Chiao Tung University, Hsinchu, Taiwan}

\begin{abstract}
Existence and stability analysis of the Kantowski-Sachs type
universe in a higher derivative induced gravity theory is studied
in details. Existence of one stable mode and one unstable mode is
shown to be in favor of the inflationary universe. As a result,
the de Sitter background can be made to be stable against
anisotropic perturbations with proper constraints imposed on the
coupling constants of the induced gravity model.
\end{abstract}

\maketitle

\section{Introduction}

Inflationary theory provides a natural resolution for the the
flatness, monopole, and horizon problems of our present universe
described by the standard big bang cosmology \cite{inflation}. In
particular, our universe is homogeneous and isotropic to a very
high degree of precision \cite{data,cobe}. Such an universe can be
described by the well known Friedmann-Robertson-Walker (FRW)
metric \cite{book}.

Moreover, gravitational physics could be different from the
standard Einstein models near the Planck scale
\cite{string,scale}. For example, quantum gravity or string
corrections could lead to some interesting cosmological
applicastions\cite{string}. In particular, investigations have
been conducted on the possibility of deriving inflation from
higher order gravitational corrections
\cite{jb1,jb2,Kanti99,dm95}.

For example, a general analysis on the stability conditions of
gravity theories could be useful in screening physical models
compatible with our physical universe. In particular, the
stability condition for a variety of pure gravity theories as a
potential candidate of inflationary universe in the flat
Friedmann-Robertson-Walker (FRW) space is derived in Ref.
\cite{dm95,kp91,kpz99}.

In addition, the highly isotropic universe should evolve from some
initially anisotropic state before it becomes isotropic to such a
high degree of precision. Nonetheless, it is interesting by itself
to study the stability analysis of the anisotropic space during
the post-inflationary epoch even anisotropy can be smoothed out by
the proposed inflationary process. One would like to know whether
our universe can evolve from certain anisotropic universe to a
stable and isotropic final state. In particular, it is known that
such inflationary solution exists for an NS-NS model with a
metric, a dilaton, and an axion field \cite{CHM01}. This
inflationary solution is also shown to be stable against small
perturbations \cite{ck01}. Note that similar stability analysis
has also been studied for a various of interesting
models\cite{kim,abel}.

The importance of higher derivative models have been known in many
aspects. In particular, higher derivative terms are known to be
important for the Planck scale physics\cite{kim,dm95}. For
example, higher order corrections from quantum gravity or string
theory have been considered as possible inflationary models
\cite{green}. In addition, higher derivative terms also arise as
the quantum corrections of the matter fields \cite{green}.
Therefore, it is important to study the implications of the
stability analysis of all possible higher derivative models.

Recently, there are also growing interests in the study of
Kantowski-Sachs (KS) type anisotropic spaces\cite{BPN,LH,NO}. We
will hence try to study the problem of existence and stability
conditions of an inflationary de Sitter final state for some
higher derivative model in Kantowski-Sachs spaces. In particular,
a large class of pure gravity models with inflationary KS$/$FRW
solutions was presented in Ref. \cite{kao06}. Any KS type solution
leading itself to an asymptotic FRW final state will be referred
to as the KS$/$FRW solution in this paper for convenience.

It has been shown that the existence of a stable de Sitter
background is closely related to the choices of the coupling
constants. Indeed, a pure gravity model given below: \cite{kao06}
\begin{equation}
{\cal L} = -R - \alpha R^2 - \beta R^\mu_\nu R^\nu_\mu + \gamma
R^{\mu \nu}_{\;\;\; \beta \gamma} \, R^{\beta \gamma}_{\;\;\;
\sigma \rho} \, R^{\sigma \rho}_{\;\;\; \mu \nu}
\end{equation}
admits an inflationary solution with a constant Hubble parameter
determined by $H_0^4 = 1/4 \gamma$ if $\gamma
>0$. Here $\alpha$, $\beta$, and $\gamma$ are coupling constants.
This shows that (a) the $\gamma$ $R$-cubic term determines
the scale of the inflation characterized by the Hubble parameter
$H_0$ and (b) quadratic terms are irrelevant to the scale $H_0$ in
the de Sitter phase. The quadratic terms are, however, important
to the stability of the de Sitter phase.

Indeed, perturbing the KS type metric with $H_i \to H_0 +\delta
H_i$, one can show that
\begin{equation}
\delta H_i = c_i \exp [{-3H_0t \over 2} (1+ \delta_1) ] + d_i \exp
[{-3H_0t \over 2} (1- \delta_1) ]
\end{equation}
for
\begin{equation}
\delta_1 = \sqrt{1+ 8/[27-9 (6 \alpha+2 \beta)H_0^2 ] }
\end{equation}
and some arbitrary constants $c_i, d_i$ to be determined by the
initial perturbations. Here $H_i\equiv \dot{a}_i/a_i$ with
$a_i(t)$ the scale factor in $i$-direction. We will describe the
notation shortly in section II. It is easy to see that any small
perturbation $\delta H_i$ will be stable against the de Sitter
background if both modes characterized by the exponents
\begin{equation}
\Delta_{\pm} \equiv - [3H_0t / 2] [1 \pm \delta_1]
\end{equation}
are all negative. This will happen if $\delta_1 < 1$. In such
case, the inflationary de Sitter space will remain a stable
background as the universe evolves. More specifically, the
constraint $ \beta H_0^2 > 35/18$ is required for both modes to be
stable.

It can be shown that the stability equation for the anisotropic KS
space and the stability equation for the isotropic FRW space in
the presence of the same inflationary de Sitter background turns
out to be identical \cite{dm95,kp91, kpz99}. Therefore, the
stability of isotropic perturbations also ensures the stability of
the anisotropic perturbations. The stability of the isotropic
perturbations for the FRW space is important for any physical
models. Unfortunately, inflationary models that are stable against
any isotropic perturbations will have problem with the graceful
exit process. Therefore, the pure gravity model may have troubles
dealing with the stability and exit mechanism all together.

Instead of the pure gravity theory, a slow rollover scalar field
may help resolving this problem. An inflationary de Sitter
solution in a scalar-tensor model is expected to have one stable
mode (against the perturbation in $\delta H_i$ direction) and one
unstable mode (against the perturbation in $\delta \phi$
direction). As a result, the inflationary era will come to an end
once the unstable mode takes over after a brief period of
inflationary expansion. Therefore, we propose to study the effect
of such theory.

In particular, we will show in this paper that the roles played by
the higher derivative terms are dramatically different in the
inflationary phase of our physical universe in both pure gravity
theory and scalar-tensor theory. First of all, third order term
will be shown to determine the expansion rate $H_0$ for the
inflationary de Sitter space. The quadratic terms will be shown to
have nothing to do with the expansion rate of the background de
Sitter space. They will however affect the stability condition of
the de Sitter phase. Their roles played in the existence and
stability condition of the evolution of the de Sitter space are
dramatically different.

\section{Non-redundant field equation and Bianchi identity in KS space}

Given the metric of the following form:
\begin{equation}
ds^2=- dt^2+c^2(t)dr^2 + a^2(t) ( d^2 \theta +f^2(\theta) d
\varphi^2)
\end{equation}
with $f(\theta)= (\theta, \sinh \theta, \sin \theta)$ denoting the
flat, open and close anisotropic space known as Kantowski-Sachs
type anisotropic spaces. More specifically, Bianchi I (BI),
III(BIII), and Kantowski-Sachs (KS) space corresponds to the flat,
open and closed model respectively. This metric can be rewritten
as
\begin{equation} \label{metric}
ds^2=- dt^2+a^2(t)({dr^2\over {1-kr^2}}+r^2d\theta ^2) + a_z^2(t)
dz^2
\end{equation}
with $r$, $\theta$, and $z$ read as the polar coordinates and $z$
coordinate for convenience and for easier comparison with the FRW
metric. Note that $k=0,1,-1$ stands for the flat, open and closed
universes similar to the FRW space.

Writing $H_{\mu \nu} \equiv G_{\mu \nu} -T_{\mu \nu}$, Einstein
equation can be written as $D_\mu H^{\mu \nu}=0$ incorporating the
Bianchi identity $D_\mu G^{\mu \nu}=0$ and the energy momentum
conservation $D_\mu T^{\mu \nu}=0$. Here $G^{\mu \nu}$ and $T^{\mu
\nu}$ represent the Einstein tensor and the energy momentum tensor
coupled to the system respectively. With the metric
(\ref{metric}), it can be shown that the $r$ component of the
equation $D_\mu H^{\mu \nu}=0$ implies that
\begin{equation}
H^r_{\; r}=H^\theta_{\;\theta}.
\end{equation}
This result also says that any matter coupled to the system has
the symmetric property $T^r_{\;r}=T^\theta_{\;\theta}$. In
addition, the equations $D_\mu H^{\mu \theta}=0$ and $D_\mu H^{\mu
z}=0$ both vanish identically for all kinds of energy momentum
tensors. More interesting information comes from the $t$ component
of this equation. It says:
\begin{equation}
(\partial_t + 3 H) H^t_{\; t} = 2 H_1 H^r_{\; r} +H_zH^z_{\; z}.
\end{equation}
This equation implies that (i) $H^t_{\; t}=0$ implies that
$H^r_{\; r}=H^z_{\; z}=0$ and (ii) $H^r_{\; r}=H^z_{\; z}=0$ only
implies $(\partial_t + 3 H) H^t_{\; t} =0$ instead of $H^t_t=0$.
Case (ii) can be solved to give $ H^t_{\; t} =$ constant $\times
\exp[-a^2a_z]$ which approaches zero when $a^2a_z \to \infty$. For
the anisotropic KS spaces, the metric contains two independent
variables $a$ and $a_z$. The Einstein field equations have,
however, three non-vanishing components: $H^t_{\; t} =0$, $H^r_{\;
r}=H^\theta_{\;\theta} =0$ and $H^z_{\; z} =0$. The Bianchi
identity implies that the $tt$ component is not redundant and will
hence be retained for complete analysis. Ignoring either one of
the $rr$ or $zz$ components will not affect the final result of
the system. In short, the $H^t_t=0$ equation, known as the
generalized Friedman equation, is a non-redundant field equation
as compared to the $H^r_r=0$ and $H^z_z=0$ equations.

In addition, restoring the $g_{tt}$ component $b^2(t)=1/B_1$ will
be helpful in deriving the non-redundant field equation associated
with $G_{tt}$ that will be shown shortly. More specifically, the
generalized KS metric will be written as:
\begin{equation}  \label{metricb}
ds^2=-b^2(t) dt^2+a^2(t)({dr^2\over {1-kr^2}}+r^2d\theta ^2) +
a_z^2(t) dz^2.
\end{equation}
In principle, the Lagrangian of the system can be reduced from a
functional of the metric $g_{\mu \nu}$, ${\cal L}(g_{\mu \nu})$,
to a simpler function of $a(t)$ and $a_z(t)$, namely $L(t) \equiv
a^2 a_z {\cal L}(g_{\mu \nu}(a(t), a_z(t)))$. The equation of
motion should be reconstructed from the variation of the reduced
Lagrangian $L(t)$ with respect to the variable $a$ and $a_z$. The
result is, however, incomplete because, the variation of $a$ and
$a_z$ are related to the variation of $g_{rr}$ and $g_{zz}$
respectively. The field equation from varying $g_{tt}$ can not be
derived without restoring the variable $b(t)$ in advance. This is
the motivation to introduce the metric (\ref{metricb}) such that
the reduced Lagrangian $L(t) \equiv ba^2 a_z {\cal L}(g_{\mu
\nu}(b(t),a(t), a_z(t)))$ retains the non-redundant information of
the $H^t_t=0$ equation. Non-redundant Friedmann equation can be
reproduced resetting $b=1$ after the variation of $b(t)$ has been
done.

After some algebra, all non-vanishing components of the curvature
tensor can be computed: \cite{kao06}
\begin{eqnarray}
R^{ti}_{\;\;\;tj} &=& [{1\over 2}\dot{B}_1H_i+B_1 ( \dot{H}_i+H^2_i) ] \delta^i_j ,\\
R^{ij}_{\;\;\;kl} &=& B_1H_iH_j \; \epsilon^{ijm}\epsilon_{klm}+{k
\over a^2} \epsilon^{ijz}\epsilon_{klz}
\end{eqnarray}
with $H_i \equiv ( \dot{a} /a, \dot{a} /a, \dot{a_z} /a_z)$
$\equiv (H_1,H_2=H_1, H_z)$ for $r, \theta$, and $z$ component
respectively.

 Given a Lagrangian
$L = \sqrt{g} {\cal L}=L(b(t), a((t), a_z(t))$, it can be shown
that
\begin{eqnarray}
L &=& { a^2 a_z \over \sqrt{B_1}} {\cal L} (R^{ti}_{\;\;\;tj},
R^{ij}_{\;\;\;kl}) = { a^2 a_z \over \sqrt{B_1}} {\cal L} (H_i,
\dot{H}_i, a^2)
\end{eqnarray}
The variational equations for this action can be shown to be:
\cite{kao06}
\begin{eqnarray} \label{key0}
{\cal L} +H_i ( {d \over dt} +3H )L^i &=& H_iL_i + \dot{H}_i L^i \\
{\cal L} + ( {d \over dt} +3H )^2 L^z &=& ( {d \over dt} +3H )L_z
\end{eqnarray}
Here $L_i \equiv \delta {\cal L} /\delta H_i$,  $L^i \equiv \delta
{\cal L} /\delta {\dot H}_i$, and $3H \equiv \sum_i H_i$. For
simplicity, we will write ${\cal L}$ as $L$ from now on in this
paper. As a result, the field equations can be written in a more
comprehensive form:
\begin{eqnarray} \label{key}
DL &\equiv&  L +H_i ( {d \over dt} +3H )L^i - H_iL_i - \dot{H}_i L^i =0\\
D_z L &\equiv& L + ( {d \over dt} +3H )^2 L^z - ( {d \over dt} +3H
)L_z =0  \label{zeq}
\end{eqnarray}

\section{higher derivative induced gravity model}

In this section, we will study the higher derivative induced
gravity model:
\begin{equation}
{ L} = -{\epsilon \over 2} \phi^2 R - \alpha R^2 - \beta R^\mu_\nu
R^\nu_\mu +  {\gamma \over \phi^2} R^{\mu \nu}_{\;\;\; \beta
\gamma} \, R^{\beta \gamma}_{\;\;\; \sigma \rho} \, R^{\sigma
\rho}_{\;\;\; \mu \nu} -{1 \over 2} \partial_\mu \phi \partial^\mu
\phi - V(\phi) \equiv {\epsilon \over 2} \phi^2  L_0 + L_2 +
{\gamma \over \phi^2} L_3 +L_\phi
\end{equation}
with $L_0 = -R$, $L_2 =-\alpha R^2 - \beta R^\mu_\nu R^\nu_\mu$,
$L_3=  R^{\mu \nu}_{\;\;\; \beta \gamma} \, R^{\beta
\gamma}_{\;\;\; \sigma \rho} \, R^{\sigma \rho}_{\;\;\; \mu \nu}$
and $ L_\phi = -{1 \over 2} \partial_\mu \phi \partial^\mu \phi -
V(\phi)$ denoting the lowest order curvature coupling, the higher
order terms, and the scalar field Lagrangian respectively. Induced
gravity models assume that all dimensionful parameters or coupling
constants are induced by a proper choice of dynamical fields. For
example, the gravitational constant is replaced by $8\pi G
=2/(\epsilon \phi^2)$ as a dynamical field. In addition, the
cosmological constant becomes $V(\phi)$ in this model. There is no
need for any induced parameters for the quadratic terms $R^2$ and
$R_{\mu \nu}^2$ because the coupling constants $\alpha$ and
$\beta$ are both dimensionless by itself. This action of the
system is also invariant under the global scale transformation:
$g_{\mu \nu} \to \Lambda^{-2} g_{\mu \nu}$ and $\phi \to \Lambda
\phi$ with arbitrary constant parameter $\Lambda$.

The corresponding Lagrangian can be shown to be:
\begin{eqnarray}&&
L= \epsilon \phi^2(2A+B+2C+D)- 4 \alpha \left[
4A^2+B^2+4C^2+D^2+4AB+8AC+4AD+4BC+2BD+4CD \right] \nonumber
\\ && - 2 \beta \left[3A^2+B^2+3C^2+D^2+2AB+2AC+2AD+2BC+2CD
\right]+ +8{\gamma \over \phi^2} \left[2A^3+B^3+2C^3+D^3
\right]\nonumber \\
&&+{1 \over 2}\dot{\phi}^2-V(\phi)
\end{eqnarray}
with
\begin{eqnarray}
&& A=\dot{H}_1+H_1^2, \\
&& B= H_1^2 + {k \over a^2}, \\
&& C=H_1H_z, \\
&& D=\dot{H}_z+H_z^2 .
\end{eqnarray}
This Lagrangian can be shown to reproduce the de Sitter models
when we set $H_i \to H_0$ in the isotropic limit. The Friedmann
equation reads:
\begin{eqnarray} \label{keyphi}
{1 \over 2} \epsilon \phi^2 DL_0 +DL_2 + {\gamma \over \phi^2}
DL_3 + \epsilon \phi \dot{\phi} H_i {L_0^i} - 2 {\gamma \over
\phi^3} \dot{\phi} H_i {L_3^i} = {1 \over 2}{\dot{\phi}}^2
+V(\phi)
\end{eqnarray}
for the induced gravity model. In addition, the scalar field
equation can be shown to be:
\begin{equation} \label{phieq}
\ddot{\phi} +3 H_0 \dot{\phi} +V'= \epsilon \phi L_0 - 2 {\gamma
\over \phi^3} L_3.
\end{equation}
The leading order de Sitter solution with $\phi=\phi_0$ and
$H_i=H_0$ for all directions can be shown to be:
\begin{eqnarray}
V_0 &\equiv& V(\phi_0) = 3 \epsilon' \phi_0^2 H_0^2 ,\\
V'(\phi_0) &=& 12 \epsilon' \phi_0 H_0^2
\end{eqnarray}
under the slow roll-over assumption $\ddot{\phi} << V'$ and $H_0
\dot{\phi} << V'$. Here $\epsilon' \equiv \epsilon [1-8\gamma
H_0^4/(\epsilon \phi_0^4)]$. Indeed, It can be shown that in the
de Sitter inflationary phase, the ignored part of the scalar field
equation evolves as $\ddot{\phi} +3 H_0 \dot{\phi} \sim 0$. This
equation leads to the approximated solution
\begin{equation}
\phi \sim \phi_0 + {\dot{\phi}_0 \over 3 H_0}[1 - \exp (- 3 H_0 t)
] \label{phit}
\end{equation}
during the de Sitter phase $H_i=H_0$. This result is clearly
consistent with the slow roll-over assumption we just made. In
summary, the leading order equations give us a few constraints on
the field parameters:
\begin{equation} \label{constraint}
4V_0 =\phi_0 {\partial V \over \partial \phi} (\phi= \phi_0) =12
\epsilon' \phi_0^2 H_0^2 .
\end{equation}
An appropriate effective spontaneously symmetry breaking potential
$V$ of the following form
\begin{equation} \label{V0}
V(\phi)= { \lambda \over 4} (\phi^2-\phi_0)^2 +6\epsilon' H_0^2
(\phi^2-\phi_0^2) + 3 \epsilon' H_0^2 \phi_0^2
\end{equation}
with arbitrary coupling constant $\lambda$  can be shown to be a
good candidate satisfying all the scaling conditions
(\ref{constraint}).

The value of $H_0$ can be chosen to induce enough inflation for a
brief moment as long as the slow rollover scalar field remains
close to the initial state $\phi=\phi_0$. The de Sitter phase will
hence remain valid and drive the inflationary process for a brief
moment determined by the decaying speed of the scalar field. The
stability conditions associated with this effective potential
implied by the field equations will be studied in the following
section.

Note that the local extremum of this effective potential can be
shown to be $\phi=0$ (local maximum) and $\phi^2=\phi_m^2=
\phi_0^2- 12 \epsilon' H_0^2/\lambda <\phi_0^2$ (local minimum).
In addition the minimum value of the effective potential can be
shown to be
\begin{equation}
V_m =V_0 -36 \epsilon'^2 H_0^4/\lambda <V_0.
\end{equation}
The constraint $V_m>0$ implies that $\lambda \phi_0^2 > 12
\epsilon' H_0^2$. Or equivalently, it implies that $\phi_m^2 >0$.
In addition, we will set $\epsilon \phi_m^2/2=1/(8\pi G)=1$ in
Planck unit later in this paper.

When the scalar field settles down to the local minimum $\phi_m$
of the effective potential at large time in the post inflationary
era, it will oscillate around the local minimum and kick off the
reheating process. The scalar field will eventually become a
constant background field with a small cosmological constant $V_m
= V(\phi_m)$.

\section{Stability of higher derivative inflationary solution}

Our universe could start out anisotropic and evolves to the
present highly isotropic state in the post inflationary era.
Therefore, a stable KS$/$FRW solution is necessary for any
physical model of our universe.

Given an effective action of the sort described by Eq. (\ref{V0}),
the scaling constraints (\ref{constraint}) is required for the
existence of the de Sitter solution $H_i=H_0$ and the static
condition $\phi=\phi_0$. In addition to these constraints, small
perturbations, $H_i=H_{i0}+ \delta H_i$ and $\phi=\phi_0+\delta
\phi$, against the background de Sitter solution $(H_{i0},
\phi_0)$ may also put a few more constraint to the stability
requirement of the system. This perturbation will enable one to
understand whether the background solution is stable or not. In
particular, it is interesting to learn whether a KS $\to$ FRW
(KS/FRW) type evolutionary solution is stable or not.

It can be shown that the perturbation equation for the Bianchi
models are identical to the perturbation equation for the FRW
models. Therefore, any inflationary solutions with a stable mode
and an unstable mode will provide a natural way for the
inflationary universe to exit the inflationary phase. Such models
will, however, also be unstable against the anisotropic
perturbations. Therefore, any inflationary solutions with a stable
mode and an unstable mode is also a negative result to our search
for a stable and isotropic inflationary model. As a result, such
solution will be harmful for an anisotropic space to reach an
isotropic FRW space once the de Sitter phase start to collapse. It
will be shown shortly that the higher derivative induced gravity
theory could hopefully resolve this problem all together.

In practice, perturbing the background de Sitter solution along
the $\delta H_i$ direction should be stable for at least a brief
moment of the order $\Delta T \sim  60 H_0^{-1}$ for a physical
model. As a result around $60$ e-fold inflation can be induced
before the de Sitter phase collapses. And the resulting universe
is stable against isotropic and anisotropic perturbations. In
addition, the scalar field is expected to roll slowly from the
initial state $\phi=\phi_0$. Therefore, the perturbation along the
$\delta \phi$ direction is expected to be unstable. This will
favor the system for a natural mechanism for graceful exit. Hence,
we will try to study the stability equations of the system for
small perturbations against the de Sitter background solutions.

The first order perturbation equation for $DL$, with $H_i \to
H_0+\delta H_i$, can be shown to be:
\begin{eqnarray}\label{stable0}
\delta ( DL) &=& <H_i L^{ij} \delta  \ddot{H}_j> +3H <H_i L^{ij}
\delta \dot{H}_j> + \delta <H_i\dot{L}^i>+3H <(H_i L^i_j + L^j)
\delta
H_j> \nonumber \\
&& +<H_iL^i> \delta (3H) -<H_iL_{ij} \delta H_j>\label{stable1}
\end{eqnarray}
for any $DL$ defined by Eq. (\ref{key}) with all functions of
$H_i$ evaluated at some FRW background with $H_i=H_0$. The
notation $< A_iB_i> \equiv \sum_{i=1,z} A_iB_i$ denotes the
summation over $i=1$ and $z$ for repeated indices. Note that we
have absorbed the information of $i=2$ into $i=1$ since they
contributes equally to the field equations in the KS type spaces.
In addition, $L^{i}_{j} \equiv \delta^2 L / \delta \dot{H}_{i}
\delta H_j$ and similarly for $L_{ij}$ and $L^{ij}$ with upper
index $^i$ and lower index $_j$ denoting variation with respect to
$\dot{H}_i$ and $H_j$ respectively for convenience. In addition,
perturbing Eq. (\ref{zeq}) can also be shown to reproduce the Eq.
(\ref{stable0}) in the de Sitter phase\cite{inflation}.

In addition, it can be shown that
\begin{eqnarray}
<H_i L^{i1}> &=& 2 <H_i L^{iz}  > ,\\
<H_i L^i_1> &=& 2 <H_i L^i_z>, \\
L^1 &=& 2 L^z ,\\
<H_iL_{i1}> &=& 2 <H_iL_{iz}>,
\end{eqnarray}
in the inflationary de Sitter background with $H_0=$ constant.
Therefore, the stability equations (\ref{stable1}) can be greatly
simplified. For convenience, we will define the operator ${\cal
D}_L$ as
\begin{eqnarray}
{\cal D}_L \delta H \equiv  <H_i L^{i1}> \delta  \ddot{H} +3H <H_i
L^{i1}> \delta \dot{H} +3H <H_i L^i_1 + L^1>  \delta H+2 <H_iL^i>
\delta H - <H_iL_{i1}> \delta H=0.
\end{eqnarray}
As a result, the stability equation (\ref{stable1}) becomes
\begin{equation}
\delta (DL)=  {\cal D}_L (\delta H_1+ \delta H_z/2)={3 \over
2}{\cal D}_L (\delta H) =0
\end{equation}
with $H=(2 H_1+H_z)/3$ as the average of $H_i$.

Hence the leading order perturbation equation in $\delta H$ and
$\delta \phi$ for the Friedmann equation of this model can be
shown to be:
\begin{equation} {\gamma
H_0^4 \over \epsilon \phi_0^4} -6\epsilon (1- 24 {\gamma H_0^4
\over \epsilon \phi_0^4}) \phi_0 H_0 [\delta \dot{\phi} -H_0
\delta \phi] = {\epsilon \over 2} \phi_0^2 \delta (DL_0) +\delta
(DL_2)+ {\gamma \over \phi^2} \delta (DL_3)= {3 \over 4} \epsilon
\phi_0^2 {\cal D}_0 \delta H +{3 \over 2}{\cal D}_2 \delta H +
{3\gamma \over 2 \phi^2} {\cal D}_3 \delta H
\end{equation}
with ${\cal D}_0 \delta H \equiv {\cal D}_{L_0} \delta H $, ${\cal
D}_2 \delta H \equiv {\cal D}_{L_2} \delta H $ and ${\cal D}_3
\delta H \equiv {\cal D}_{L_3} \delta H $ as short-handed
notations. This equation can further be shown to be:
\begin{equation}
\epsilon (1- 24 {\gamma H_0^4 \over \epsilon \phi_0^4})  \phi_0
[\delta \dot{\phi} -H_0 \delta \phi] = 4( 3\alpha +\beta -6 \gamma
{H_0^2 \over \phi_0^2}) (\delta \ddot{H} +3H_0 \delta \dot{H}) +
(24 \gamma {H_0^4 \over \phi_0^2}-\epsilon \phi_0^2) \delta H.
\end{equation}
Similarly, the leading perturbation of the scalar field equation
can be shown to be:
\begin{equation}
\delta \ddot{\phi} +3H_0 \delta \dot{\phi} +(V'' -12 \epsilon'
H_0^2 - 384 \gamma {H_0^6 \over \phi_0^4})\delta \phi = 6
\epsilon' \phi_0 (\delta \dot{H} +4 H_0 \delta H)
\end{equation}
The variational equation of $a_z$ can be shown explicitly to be
redundant in the limit $H_i =H_0+\delta H_i$ and $\phi=\phi_0+
\delta \phi$ following the Bianchi identity.

Assuming that $\delta H=\exp[hH_0t] \delta H_0$ and $\delta \phi
=\exp[pH_0t] \delta \phi_0$ for some constants $h$ and $p$, one
can write above equations as:
\begin{eqnarray}
\epsilon (1- 24 {\gamma H_0^4 \over \epsilon \phi_0^4})\phi_0[p
-1] \delta \phi &=& 4( 3\alpha +\beta -6 \gamma {H_0^2 \over
\phi_0^2}) H_0 [ h^2 +3h + {24 \gamma {H_0^4 / \phi_0^2}-\epsilon
\phi_0^2 \over 4(
3\alpha +\beta -6 \gamma {H_0^2 / \phi_0^2})} ]\delta H, \\
H_0\left [ p^2+3p+{V'' \over H_0^2} -12\epsilon' -384 \gamma{
H_0^4 \over  \phi_0^4} \right ] \delta \phi &=& 6 \epsilon' \phi_0
[ h +4 ] \delta H.
\end{eqnarray}
These equations are consistent when all coefficients vanish
simultaneously. This implies that $h=-4$ and $p=1$. This set of
solution $(h,p)=(-4,1)$ hence imposes two additional constraint
\begin{equation} \label{ch}
\epsilon-16 (3 \alpha +\beta) {H_0^2 \over \phi_0^2} + 72 \gamma
{H_0^4 \over \phi_0^4}  =0,
\end{equation}
\begin{equation} \label{cp}
\lambda = 192 \gamma {H_0^6 \over \phi_0^6}  - 2 {H_0^2 \over
\phi_0^2}
\end{equation}
with $2 \lambda \phi_0^2= V_0''-12 \epsilon'H_0^2$.

The coupling constant $\lambda$ has to positive in order for the
effective potential $V(\phi)$ to be free from run-away negative
global minimum at $\phi \to \infty$. As a result, the constraints
$\epsilon>0$ and $\lambda >0$ imply that
\begin{equation} \label{gamma}
{2(3 \alpha+\beta)\phi_0^2 \over 9 H_0^2} > \gamma > {\phi_0^4
\over 96 H_0^4}.
\end{equation}
Therefore, the inflationary phase will remain stable against small
perturbation along the $\delta H(= \exp[-4H_0t] \delta H_0)$
direction. In addition, the inflationary phase also has an
unstable mode when we perturb the system along the $\delta \phi
(=\exp[H_0t] \delta \phi_0)$ direction. The unstable mode is in
fact consistent with the slow rollover assumption. The scalar
field is expected to roll slowly off the initial state $\phi_0$
for a brief moment during the inflationary era. This unstable mode
is hence responsible for the graceful exit process.

Hence such system with a stable mode and an unstable mode is a
very nice candidate for a inflationary model. The stable mode
$\delta H = \exp[-4H_0t] \delta H_0$ implies that the de Sitter
background solution will remain stable against small isotropic
perturbation. It also implies that the system is stable against
any anisotropic perturbation along all $\delta H_i$ directions.
Therefore, the induced gravity model with a scalar field can
indeed resolve the stability problem of the pure gravity model.

\section{conclusion}

The existence of a stable de Sitter background is closely related
to the choices of the coupling constants. The pure higher
derivative gravity model with quadratic and cubic interactions
\cite{kao06} admits an inflationary solution with a constant
Hubble parameter. Proper choices of the coupling constants allow
the de Sitter phase to admit one stable mode and one unstable mode
for the anisotropic perturbation.

The stable mode favors a strong inflationary period and the
unstable mode provides a natural mechanism for the graceful exit
process. It is also found that the perturbation against the
isotropic FRW background space and the perturbation against the
anisotropic KS type background space obey the same perturbation
equations. This is true for both pure and induced gravity models.
As a result, the unstable mode in pure gravity model also means
that the isotropic de Sitter background is unstable against
anisotropic perturbations. Therefore, small anisotropic could be
generated during the de Sitter phase for pure gravity model.

We have shown that, for induced gravity models, stable mode for
perturbations along the anisotropic $\delta H_i$ directions does
exist with proper constraints imposed on the coupling constants.
In addition, another unstable mode for perturbation against the
scalar field background $\phi_0$ also exists with proper
constraints. Therefore, isotropy of the de Sitter background can
remain stable during the inflationary process for the induced
gravity models.

Explicit model with a spontaneously symmetry breaking $\phi^4$
potential is presented as an example. Proper constraints are
derived for reference. It is shown that the scaling conditions
(\ref{constraint}) must hold for the initial de Sitter phase
solutions in order for the system to admit a constant background
solution during the inflationary phase. Additional constraints
(\ref{ch}) and (\ref{cp}) are also shown explicitly for this
model. In addition, the inequality shown in Eq. (\ref{gamma}) is
expected to hold for a physical model with this potential.

In summary, we have shown that a stable mode for (an)isotropic
perturbation against the de Sitter background does exist for
induced gravity model The problem of graceful exit can rely on the
unstable mode for the scalar field perturbation against the
constant phase $\phi_0$.

It is also found that the quadratic terms will not affect the the
inflationary solution characterized by the Hubble parameter $H_0$.
These quadratic terms play, however, critical role in the
stability of the de Sitter background. In addition, it is also
interesting to find that their coupling constants $\alpha$ and
$\beta$ always show up as a linear combination of $3\alpha +
\beta$ in these stability equations. Implications of these
constraints deserve more attention for the applications to the
inflationary models.

\section*{Acknowledgments}

This work is supported in part by the National Science Council of
Taiwan.

\end{document}